# The Thermodynamics of a Single Bead in a Vibrating Container

## P. Evesque

Lab MSSMat, UMR 8579 CNRS, Ecole Centrale Paris
92295 CHATENAY-MALABRY, France, e-mail: evesque@mssmat.ecp.fr

**Abstract:**
*Statistics of the 1d dynamics of a single particle in a box of length L is studied under different conditions of periodic excitation: vibrated container, or 1 wall moving periodically or two walls vibrating in opposite phases. Theoretical predictions for the mean typical speed $<v>$ are derived using some Random Phase Approximation (RPA) for small $b/L$ ratio, as functions of the amplitude $b$ and frequency $f = \nu = 2\pi/\omega$ of vibration and of the collision restitution coefficient $r = v_{out}/v_{in}$. They are compared to numerical simulations. It is concluded that RPA is valid for $b/L > 0.005$, that long time memory, i.e. non ergodicity, and/or resonance develop at large $r$ ($r > 0.95$) and/or at large $b/L$, that $<v>$ scales always as $f$, and scales as $b$ except for large $b/L$ values, i.e. $b/L > 0.02$, and that the relative velocity $<V> = [<v>/(b\omega)] \approx 1$ when $r < 0.4$-$0.5$. Relative standard deviation $\Delta V/V$ is found to be approximately constant, i.e. $\Delta V/V \approx 0.3$ except when resonance; in this case, memory effect and non ergodicity effects are observed and become too important.*

**Pacs # :** 5.40 ; 45.70 ; 64 ; 83.70.Fn

______________________________________________________________________

A great deal of effort has been sustained to understand the general problem of the mechanical behaviour of granular material under vibrational excitation [1-3]. More recently, an experiment has been performed in a Mini-Texus sounding rocket in order to determine the properties of a very loose ensemble of grains under vibration excitation and in micro-gravity conditions [4,5]. The aim of the experiment was to observe new phenomena which result from inelastic collisions and to study the properties of a "granular gas" which may be different from the ones of a classical gas for which no dissipation occurs during collisions. These sounding-rocket experiments have allowed to identify two distinct regimes: the first regime corresponds to a so dilute case that the granular gas is in some kind of Knudsen regime, the second one occurs as soon as the density is larger.

In the very dilute case, it has been shown that the pressure of the granular gas scales like the 3/2 power of the vibration velocity. When the density of the medium is increased, it has been observed for the first time that an ensemble of solid particles in erratic motion interacting only through inelastic collisions generates the formation of a dense cluster. This cluster seems to be maintained rather motionless far from the vibrating walls thanks to the existence of a "gaseous" granular phase which screens the moving walls. In the present case, this occurs as soon as the mean free path $l_c$ between two grain-grain collisions becomes smaller than the cell size L. Gaseous conditions are then only observed for a system of solids particles in the so-called Knudsen regime, $l_c > L$. These experiments tend also to demonstrate that the statistics of grain velocity does not obey the classical Boltzmann distribution even under such restrictive conditions. At last, it has been observed during these experiments that the typical speed of the grains in the gaseous phase seems to be less than the maximum speed of the walls [5] because the grain density at the wall vanishes periodically when the amplitude of vibration is large enough to allow its measure; this tends to indicate that granular gas is subject to some kind of "supersonic excitation".

These results have been confirmed and improved more recently using the Micro-gravity Airbus-A300 of Centre National d'Études Spatiales (CNES). In particular two different kinds of excitation have been used, which are either a vibrating box or a fixed box with a single vibrating wall.

So, in view of these results one can think that the thermodynamics of a very dilute granular gas shall be similar to the one of an "isolated" particle in a vibrating container. In order to prove this parallel, to identify the resemblance and the differences, it is needed to study the problem of the dynamics of a single particle in a vibrating box. This is the concern of this paper.

In this paper, the dynamics is always limited to the 1d problem; different hypotheses have been used for the vibrating walls: (i) the two walls vibrate in phase at the same frequency $f=\nu=\omega/(2\pi)$ and at the same amplitude b, (ii) an immobile wall and the other one vibrating, (iii) the two walls vibrating in opposite phases with the same amplitude. In all cases, the length L of the container has been assumed larger than twice the vibration amplitude (L>2b). The other case (L<<b) would have led to completely different results. In each case, the dependence of the particle velocity v, of its mean <v> and of its mean square root deviation $\delta v$ upon the restitution coefficient $r=v_{out}/v_{in}$ have been determined in units of $b\omega$, and as functions of b/L. It is also demonstrated using numerical simulations that the statistics of the dynamics of a single bead in a vibrating container may depend on the initial condition in some cases; so we have also investigated the dependence of <v> and $\delta v$ on these initial conditions; the simulations demonstrate that loss of initial-condition memory is rapid most often since it does not depends on the amplitude $\delta\phi$, *i.e.* $\delta\phi<10^{-4}$ or $2\pi/10$, of the initial phase in general; furthermore, in most cases the mean dynamics and the mean characteristics of the distribution of speed are also not sensitive to a change of the initial speed $v_o$ of the bead $2b\omega/1000<v_o<2b\omega$. However, one can observe some resonances for peculiar sets of (r, b/L, $\phi_b$) and some dependence of the results on initial conditions for some range of parameters b/L and r; in this case it implies strong memory effect and can mean also loss of ergodicity.

This paper is divided into two parts; in the first one the problem is settled theoretically and approximations developed in order to find the dependence of <v> as a function of the parameters r, b, $f=\omega/2\pi$ in the limit of small b/L. In the second part a systematic investigation of the real dynamics is performed using computer calculations. The results of section 1 are recovered when b/L<<1, but differences are also found which are described and discussed. The validity of the hypotheses made in section 1 are then discussed in view of these results. In Section 3, we conclude and sum up the findings; and we discuss the ability of the model to describe the case of a dilute granular gas in micro-gravity condition.

## 1. 1 bead in a vibrating container: the theoretical approach

Let us consider a 1d box of length L which is vibrated parallel to its length and which contains a single point-like particle in no-gravity condition. Let x be the coordinate of the particle and $x_{w\pm}$ the coordinate of the walls : $x_{w\pm}= \pm L/2+b_{w\pm} \cos(\omega t+\phi_{w\pm})$; this defines the amplitude $b_{w\pm}$, the phases $\phi_{w\pm}$ and the frequency $f=\omega/2\pi$ of vibration. Here after, we suppose always that $b_{w\pm} < L/2$ and very often that $b<<L/2$. Due to the excitation, the particle moves at constant speed $v=dx/dt$ during free flight between each hit of the moving wall. The mass of these walls are assumed to be much larger than the particle mass. We will forget the index $_{w\pm}$ labelling the wall from now.

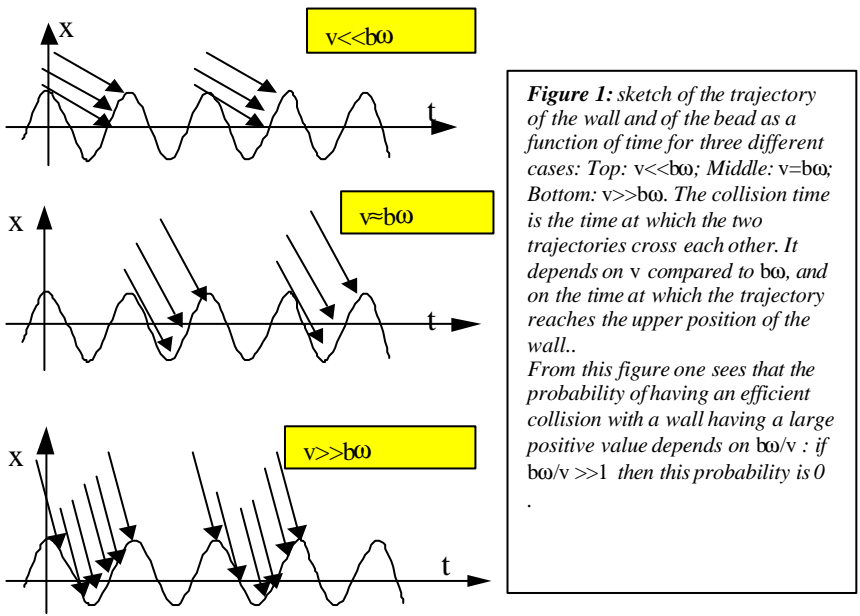

*Figure 1: sketch of the trajectory of the wall and of the bead as a function of time for three different cases: Top: $v<<b\omega$; Middle: $v\approx b\omega$; Bottom: $v>>b\omega$. The collision time is the time at which the two trajectories cross each other. It depends on v compared to $b\omega$, and on the time at which the trajectory reaches the upper position of the wall..*
*From this figure one sees that the probability of having an efficient collision with a wall having a large positive value depends on $b\omega/v$ : if $b\omega/v >>1$ then this probability is 0.*

Collisions with walls happen repetitively and are characterised by a restitution coefficient r which is defined as the ratio between the relative output velocity $v_{out}$ divided by the relative input one $v_{in}$, *i.e.* $r=-v_{out}/v_{in}$. Within such a definition, the restitution coefficient is the square root of the usual one $\varepsilon_E$ which is linked to the energy restitution: $\varepsilon_E =r^2= v^2_{out}/v^2_{in}$. $v_{out}$ and $v_{in}$ are the velocities before and after the collision in the wall frame. They are related to the bead velocity in the immobile frame by:

$$v_{out}= v_2+b\omega \sin(\omega t_{coll}) \qquad (1.a)$$

$$v_{in}= v_1+b\omega \sin(\omega t_{coll}) \qquad (1.b)$$

$$v_2 = -(1+r)b\omega \sin(\omega t_{coll}) - r\, v_1 \qquad (1.c)$$

So the bead velocity $v_2$ after the collision depends on the phase of the wall motion when collision happens. It varies from one collision to another one most likely. However it is worth noting that the maximum change of velocity $v_2-v_1$ after one collision is limited to $\pm(1+r)b\omega < \pm 2\, b\omega$ when r is large since:

$$v_2 - v_1 = \Delta v_{coll} = -(1+r)\, b\omega \sin(\omega t_{coll}) + (1-r)\, v_1 \qquad (2)$$

Now let v be the typical speed of the particle and let us assume $v \approx b\omega$ ; let us also consider the case $b \ll L$; in this case, this imposes also $b\omega/f \ll L$, which implies that the duration for the bead to move from one wall to the other one is much longer than the period $1/f$. The typical time of flight between two successive collisions is then $T_f \approx L/v$. However, the precise duration depends on the real instants of contacts with the vibrating walls. These one depends on the phase $\phi$ of the wall motion when the particle enters the interacting zone. We will define this phase as $\phi = \omega\, t_{x=\pm(L/2-b)}$, (modulo $2\pi$). It means also that the precision on $T_f$ is $1/f$ about.

Indeed, as it is sketched on Fig. 1, the situation can be quite diverse depending on the real bead speed v compared to the maximum wall speed $b\omega$. For instance, this figure shows that the wall can be hit by the bead only during given phases of the motion when $v \ll b\omega$:

♣ **If $v \ll b\omega$**, one sees that the particle can hit the walls only when the wall is located around its maximum position $\pm(L/2-b)$; furthermore, this figure shows that collisions will occur more often when the wall move back than when it moves in the direction of the particle motion; furthermore, the absolute velocity of the wall can be larger when the wall move back than when it moves down (in the figure); so, this means that the momentum transfer is strongly asymmetric, advantaging the positive transfer $\Delta v > 0$.

Let us consider a period of the wall motion which ranges from $[0,2\pi]$ and a lapse of time t which ranges from $[0,2\pi/f]$; according to Fig. 1, collisions with the wall can occur only in between the two lapses of time $[0,t_1]$ $[t_2,2\pi/f]$, with $t_1$ and $t_2$ defined as follows: time $t_1$ is the time for which the bead speed is equal to the wall speed. So $-b\omega \sin(\omega t_1) = v$ and hence:

$$t_1 = (1/\omega)\, \text{Arcsin}(-v/[b\omega]) \qquad (3.a)$$

The second time $t_2$ is the time at which a collision occurs between the wall and the bead in such a way as the bead of speed v has passed already at location $x = b\cos(\omega t_1)$ at time $t_1$. So one gets: $b\cos(\omega t_2) = b\cos(\omega t_1) - v(t_2-t_1)$. So $t_2$ is the solution which is larger than $\pi/f$ of the equation:

$$v t_2 + b\cos(\omega t_2) = \cos\{\text{Arcsin}[-v/(b\omega)]\} + (v/\omega)\, \text{Arcsin}[-v/(b\omega)] \qquad (3.b)$$

According to what precedes, one can rewrite the duration range during which collision can occur as: $[t_2, t_1 + 2\pi/f]$ over the period $[t_2, t_2 + 2\pi/f]$.

♣ **If v≈bω,** the particle can penetrate deep in the interaction zone, depending on the initial value of ϕ. The maximum momentum transfer is obtained when the wall speed is the faster in the direction opposite to the particle; this occurs when $\omega t_{coll}=3\pi/2$ , and the transfer is $(1+r)b\omega$ ; conversely, the minimum speed transfer is obtained when $\omega t_{coll}=0$ and it is a negative transfer whose value is -$(1+r)b\omega-(1-r)v_1$. So the transfer of speed can vary in a large domain and depends on the initial phase ϕ and on $v_1$. Furthermore, all events do not happen with the same probability as it will be discussed in the next subsection §-1.1.

♣ **If v>>bω,** collisions can occur at any period of the wall motion phase; non symmetry between speed transfer diminishes.

In order to go further in the theoretical description, we need to make some hypothesis. Let us state this hypothesis as follows

### *1.1. The random phase approximation RPA:*

Let us now consider that the cell is quite large compared to the amplitude of vibration b. In this case, the time required by the bead to go from one wall to the other one is much longer than a period of vibration T=1/f, *i.e.* it is a large number N of periods. Let us then handle the problem as follows:

Let us assume that we know the bead velocity within a given uncertainty δv only; if δv is large enough, then δv N T≈L (δv/v) is larger than 2b ; furthermore this situation can happen very often, since one assumes L>>b; this means that the uncertainty on the phase of the next collision can be larger than 2π even if one knows rather precisely v. So, it means that one shall consider only two distinct possibilities to solve the problem: (i) either the dynamics of the bead is such as its speed can take only a discrete set of very precise values; in this case the above proposed hypothesis δv≠0 is not valid since δv=0 and the collisions occur always at very precise values of the phase ϕ. (ii) Or the possible values of v form a continuous set; in this last case one can expect that possible values of ϕ form also a continuous set and that both sets are statistically not correlated; this is what will be supposed true in the following. In this case one can solve the problem in a statistical way using a self consistent approach:

Indeed, considering a bead whose speed v is in the range v and v+δv, this bead enters the zone of interaction with the bottom wall when its position x is x=-L/2+b; due to the uncertainty on v, this bead enters this zone with a random phase compared to the wall motion as soon as L δv/v > 2L. So, its probability to hit the wall between time t and time t+δt is just proportional to the size of the visited volume δϖ during time δt, which is in turn proportional to the relative speed between the wall $v_w$ and the bead v, since $\delta\varpi=(v-v_w)\delta t=[v+b\omega\sin(\omega t)]\delta t$. Now, be p(v) the probability that the bead has a speed v, the mean speed of the bead is:

$$<v>= \int v\, p(v)dv \qquad (4)$$

which is then the mean speed of the bead before the collision. We will assume that this mean speed does not evolve when the system evolves. This will give a self consistent relation.

## 1.2. RPA self consistent approach: The case of a moving cell:

So let us now imagine how the system evolves and let us considers $2N_b$ identical boxes of size L, the first halve $N_b$ of the 2 $N_b$ boxes corresponds to boxes with particle having a positive v, the other $N_b$ halve to boxes with particle having a negative speed. So, p(v) δv is just proportional to the number of boxes N(v)δv containing a bead of speed v within δv. At a given time this is the distribution. But after a while $\Delta t$, which will be chosen large compared to the period of the wall motion T=1/f and small compared to the time of flight L/v, the beads in the boxes have moved and the distribution has evolved; the proportion of boxes whose particles have not undergone a collision depend on the real speed v of the particle the box contains; hence it is n(v)δv (L-v$\Delta$t)/L . So, one can write:

$$N_b = \int n(v)\delta v \quad (5)$$

$$N_b \langle v \rangle = \int v\, n(v)\delta v \quad (6)$$

$$N_b \langle v \rangle = \int v\, n(v)\delta v [L-v\Delta t]/L + \int n(-v)\delta v\, [v\Delta t/L] \{\int v'\, p_{coll}(-v \to v')dv'\} \quad (7)$$

where $p_{coll}(-v \to v')$ is the probability that the particle with speed -v before the collision gets the speed v' after the collision. As the problem is symmetric, one has n(v)=n(-v). So, developing Eq. (7) as $N_b\langle v\rangle = \int v n(v)\delta v - (\Delta t/L)\int v^2 n(v)\delta v + \int n(v)\delta v [v\Delta t/L]\{\int v' p_{coll}(-v \to v')dv'\}$ and using Eq. (6) lead to the self consistent equation:

$$\int v^2\, n(v)\delta v = \int n(v)\delta v\, v\, \{\int v'\, p_{coll}(-v \to v')dv'\} \quad (8)$$

for which v' is given by Eq. (1.c); it depends on the time $t_{coll}$ at which collision occurs; in the same way, since v' depends on $t_{coll}$, $p_{coll}(-v \to v')$ depends also on the time $t_{coll}$ at which the collision occurs. So, one can express Eq. (8) as a function of $t_{coll}$:

$$\int v^2\, n(v)\delta v = \int n(v)\delta v\, v\, \left\{\int_0^T [rv+(1+r)b\omega\sin(\omega t_{coll})]\, [v+b\omega\sin(\omega t_{coll})]dt_{coll}\right\}$$

$$\Big/ \int_0^T [v+b\omega\sin(\omega t_{coll})]dt_{coll} \quad (9)$$

$$\int v^2\, n(v)\delta v = \int v^2\, n(v)\delta v \int_0^T [r+(1+r)(b\omega/v)\sin(\omega t_{coll})]\, [1+(b\omega/v)\sin(\omega t_{coll})]dt_{coll}/T$$

$$(10)$$

Last integral of this equation is just the average, performed over a period T=1/f, of the expression $[r+(1+r)(b\omega/v)\sin(\omega t_{coll})]\, [1+(b\omega/v)\sin(\omega t_{coll})]$ for which either constant terms or terms varying as $\sin^2(x)$ do not vanish. So Eq. (10) becomes:

$$\int v^2 n(v)\delta v = \int v^2 n(v)\delta v \{r+(1+r)(b\omega/v)^2/2\} \qquad (12)$$

which can be rewritten as:

$$\{1-r\}\int v^2 n(v)\delta v = (1+r)(b\omega)^2/2 \int n(v)\delta v$$

whose solution is:

$$<v^2> = (b\omega)^2 \{(1+r)/(1-r)\}/2 \qquad (13)$$

Making the approximation that the square of the mean speed $<v>^2$ is not too different from $<v^2>$ because the distribution is not too broad leads to:

$$<v> \approx (b\omega)(1+r)^{1/2}/[2(1-r)]^{1/2} \qquad (14)$$

Table 1 reports the mean speed $<V>=v/(b\omega)$ in reduced unit $(b\omega)$ predicted by the above self consistent approach when $<V>$ is larger than 1. Indeed, this approach has to be modified when $<V>$ is smaller than 1 since not all values of $t_{coll}$ can be achieved in such cases: according to Fig. 1, only those times $t_{coll}$ located in between $[t_2,T+t_1]$ are accessible during the time period $[t_2,T+t_2]$ as it has been shown previously; $t_2$ & $t_1$ are given by Eqs. (3.b & 3.a). Owing to this, Eq. (10) becomes:

$$\int v^2 n(v)\delta v = \int v^2 n(v)\delta v \int_{t_1(v)}^{t_2(v)+T}[r+(1+r)(b\omega/v)\sin(\omega t_{coll})]$$
$$[1+(b\omega/v)\sin(\omega t_{coll})]dt_{coll}/T \qquad (15)$$

This effect will not be considered here because it seems to be negligible in the present case of present excitation since $v/(b\omega)$ remains always larger than 0.7.

| r | 0.1 | 0.2 | 0.5 | 0.7 | 0.8 | 0.9 | 0.95 | 0.97 | 0.98 | 0.99 |
|---|---|---|---|---|---|---|---|---|---|---|
| **1-r** | 0.9 | 0.8 | 0.5 | 0.3 | 0.2 | 0.1 | 0.05 | 0.03 | 0.02 | 0.01 |
| v/(bω) | 0.78 | 1.2 | 1.4 | 1.8 | 2.2 | 3.1 | 4.5 | 5.7 | 7 | 10 |

***Table 1:*** *mean speed* v *of a particle contained in a vibrating* 1d *box, vibrating amplitude* b *and frequency* f=ω/2π, *box size* L, *as a function of the restitution coefficient* r= $v_{out}/v_{in}$. r= $(\varepsilon_E)^{1/2}$ *is the square root of the energetic restitution* $\varepsilon_E = (v_{out}/v_{in})^2$ *which is usually defined for collisions. It is valid in the case* b<<L.

However, the RPA cannot predict any discrete distribution of speed, nor any non ergodic behaviour; furthermore it assumes that L>>b, which limits strongly the domain of application; it supposes also that the speed v cannot tends to 0, otherwise δv/v can become quite large. All these constrains limit the domain of application of this model. This is why the next section reports the study of the same problem through computer simulations. Test of validity of Eq. (14) will be performed and the domain of application will be found as functions of the different parameters b, L, f, r. Also, slightly different problem will be studied, such as the motion of a single bead in a non moving box with either a single vibrating piston or with two pistons vibrating in opposite phase.

***Remark 1: Simplified calculation:*** At this stage, it is worth noting that Eq. (14) could have been obtained in a much simpler manner, just by considering the motion of a typical bead before and after a collision; the speed v of this typical bead shall be the mean speed before colliding and its mean velocity shall be also the mean velocity after the collision. So one gets:

$$<v> = \int_0^T [r<v> + (1+r) b\omega \sin(\omega t_{coll})] [1 + (b\omega/<v>)\sin(\omega t_{coll})] dt_{coll}/T \quad (16)$$

whose integration gives $<v>=r<v>+(1+r)(b\omega)^2/(2<v>)$; so it leads to Eqs. (12) & (14) when one assimilates $<v^2>$ to $<v>^2$ as we did previously and when $<v>$ is larger than $b\omega$. When this last condition, $<V>>>1$, is not satisfied one has to integrate this Eq. in between the limits given by Eqs. (3.b) & (3.a) as it has been proposed earlier; and identical behaviours result.

***Remark 2: Stability of the solution:*** Starting from Eq. (16), one can investigate how the system evolves dynamically in mean. Let us write the equilibrium speed $v_{eq}=(b\omega)(1+r)^{1/2}/[2(1-r)]^{1/2}$; let the system starts from a slightly perturb average: $v=v_{eq}+\delta v$; the evolution of the mean velocity obeys Eq. (16); it results from this that after one bouncing the mean velocity can be written as $v'=v_{eq}+\delta v'$, with $\delta v'$ given by applying Eq. (16) to $v_{eq}+\delta v$. This leads to:

$$v_{eq}+\delta v'=r(v_{eq}+\delta v)+[b^2\omega^2(1+r)/2]/(v_{eq}+\delta v)$$

which leads to

$$\delta v' = r \, \delta v \quad (17)$$

Indeed this result means that the dynamics converges towards the solution given by Eq. (14). And the smaller the r the faster the convergence. This last point means that dynamical oscillations can be more easily generated at large r than when r is small. This will be confirmed by the numerical simulations of next section, see for instance §-2 Figs. 2 & 3. Eq. (17) holds approximately, because it does not take into account the distribution of v, which can be broadened after a collision. It is then only an approximate behaviour which is probably only valid when r>1. Furthermore, Eq. (16) does not holds true at small r, *i.e.* r<0.1, since limits of integration of Eq. (16) shall be modified to take into account the fact that collisions can not occur at any moment, *cf.* Fig. 1. So, Eq. (17) does not apply when r is too small; the stability under such a condition, *i.e.* $<v><<b\omega$, can be studied, but will not be discussed here.

## *1.3. RPA self consistent approach: The case of 1 moving wall and 1 fix wall:*

We use now the same approach to predict the mean velocity in the case of a box closed by a moving wall and a fix wall; in this case two velocities $v_+$ and $v_-$ have to be considered, since the particle receive some kinetic energy in average from the moving wall whereas it does not from the fix wall. This means that the problem is characterised by the four following unknowns which are the two distributions $n_+$, $n_-$ of particle number and the two speeds $v_+$, $v_-$. In order to solve this problem one can write:

*Conservation of flow imposes:*

$$\int v_+ \, n_+(v_+)/L \, dv_+ = \int v_- \, n_-(v_-)/L \, dv_- \qquad (18)$$

*permanent flow imposes:*

$$\int v_+ \, n_+(v_+)(L-v_+\Delta t)/L \, dv_+ + \int [\int p(v_- \to v_+)v_+ \, dv_+] \, n_-(v_-) \, (v_-\Delta t)/L \, dv_- = \int v_+ \, n_+(v_+)/L \, dv_+ \qquad (3.16)(19)$$

and the same Equation when replacing $v_+ \leftrightarrow v_-$. Taking into account Eq. (18), Eq. (19) writes:

$$\int [\int p(v_- \to v_+)v_+ \, dv_+] \, n_-(v_-) \, (v_-\Delta t)/L \, dv_- = \int v_+ \, n_+(v_+) v_+ \Delta t/L \, dv_+ \qquad (20.a)$$

or

$$\int n_-(v_-) \, v_- \, p(v_- \to v_+) v_+ \, dv_+] \, dv_- = \int v_+^2 \, n_+(v_+) \, dv_+ \qquad (20.b)$$

Taking into account the fact that $v_\pm$ is given by Eq. (1.c) and $p(v_- \to v_+)$ by $[1+(b\omega/v_-)\sin(\omega t_{coll})]dt_{coll}/T$, one gets when $<v_->\, >> b\omega$:

$$\int v_+^2 \, n_+(v_+) \, dv_+ = \int n_-(v_-) \, v_- \, p(v_- \to v_+) v_+ \, dv_+ \, dv_- =$$

$$\int v_+^2 \, n_+(v_+) \, dv_+ = \int n_-(v_-) \, v_- \, dv_- \int_0^T [(1+r)b\omega \sin(\omega t_{coll}) + r \, v_-] \, [1+(b\omega/v_-)\sin(\omega t_{coll})]dt_{coll}/T \qquad (21)$$

$$\int v_+^2 \, n_+(v_+) \, dv_+ = \int n_-(v_-) \, v_-^2 \, dv_- \{r + (1+r)b^2\omega^2/(2v_-^2)\}$$

$$\int v_+^2 \, n_+(v_+) \, dv_+ = r \int n_-(v_-) \, v_-^2 \, dv_- + (1+r)b^2\omega^2/2 \int n_-(v_-) \, dv_- \qquad (22)$$

Eq. (22) is the self consistent equation when $<v_-> \, >> b\omega$; it has to be associated to the flow-conservation Equation:

$$\int v_+ \, n_+(v_+) \, dv_+ = \int v_- \, n_-(v_-) \, dv_- \qquad (18)$$

As previously, using the approximation consisting in neglecting the width of the distribution of the particle speed so that $<v_\pm^2>=<v_\pm>^2$, these two equations can be rewritten:

$$N_+ <v_+> = N_- <v_-> \qquad (23.a)$$

$$N_+ <v_+>^2 = r \, N_- <v_->^2 + (1+r)(b\omega)^2 \, N_-/2 \qquad (23.b)$$

And due to symmetry of the problem and to the fact that the other wall has an amplitude b=0:

$$N_- <v_->^2 = r\, N_+ <v_+>^2 \qquad (23.c)$$

Combining Eqs. (23.a) & (23.c) gives:

$$<v_-> = r <v_+> \qquad (24.a)$$

and

$$N_+ = r\, N_- \qquad (24.b)$$

$$V_- = v_-/(b\omega) = \{r/[2(1-r)]\}^{1/2} \qquad (24.c)$$

$$V_+ = v_+/(b\omega) = \{1/[2r-2r^2]\}^{1/2} \qquad (24.d)$$

As mentioned in §-1.2, Eqs. (24.c) & (24.d) do apply only when $<V_->>1$, otherwise the integral limits of Eqs. (21) & (22) shall be reduced to $t_2$ and $t_1+T$ given by Eq. (3). Table 2 gives the mean speed predicted by Eqs. (24.c) & (24.d) when $<V_->$ is larger than 1. Table 2 values such as $<V><1$ come from an other estimate, *cf.* Remark 2.

| r | 0.05 | 0.1 | 0.2 | 0.5 | 0.6 | 0.7 | 0.8 | 0.9 | 0.95 | 0.97 | 0.98 | 0.99 |
|---|---|---|---|---|---|---|---|---|---|---|---|---|
| 1-r | 0.95 | 0.9 | 0.8 | 0.5 | 0.4 | 0.3 | 0.2 | 0.1 | 0.05 | 0.03 | 0.02 | 0.01 |
| v_-/(bω) | 0.02 | 0.07 | 0.32 | 0.7 | 0.91 | 1.1 | 1.4 | 2.1 | 3.08 | 4.0 | 4.95 | 7.03 |
| V_+/(bω) | 0.31 | 0.68 | 1.44 | 1.4 | 1.51 | 1.54 | 1.8 | 2.35 | 3.24 | 4.14 | 5.05 | 7.1 |

***Table 2:*** *mean speed* v *of a particle contained in a fixed* 1d *box with 1 wall vibrating at amplitude* b *and frequency* f=ω/2π, *box size* L, *as a function of the restitution coefficient* r= $v_{out}/v_{in}$. r= $(\varepsilon_E)^{1/2}$ *is the square root of the energetic restitution* $\varepsilon_E = (v_{out}/v_{in})^2$ *which is usually defined for collisions. It is valid in the case* b<<L.

**Remark 1: Simplified calculation:** In the same way as in the last subsection (§-1.2), it is worth noting that results given by Eq. (24) could have been obtained in a much simpler manner, just by considering the motion of a typical bead before and after a round trip and to impose that the speed of this typical bead shall be equal to the typical bead speed, *i.e.* the mean speed, before the round trip and after the round trip. So one gets directly the configuration:

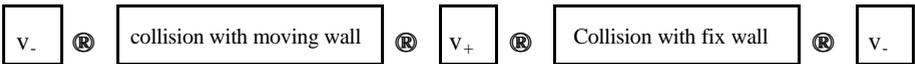

$$<v_-> = r \int_0^T [r<v_+>+(1+r)\, b\omega\, \sin(\omega t_{coll})]\, [1+(b\omega/<v_->)\sin(\omega t_{coll})]\, dt_{coll}/T \qquad (25.a)$$

$$<v_+> = \int_0^T [r<rv_+>+(1+r)\, b\omega\, \sin(\omega t_{coll})]\, [1+(b\omega/<rv_+>)\sin(\omega t_{coll})]\, dt_{coll}/T \qquad (25.b)$$

which leads to Eqs. (24.c) & (24.d). In the same way as in the previous case, complete integration is valid only when the mean velocity $<v_->=r<v_+>$ is larger than bω. In the other case, one has to change the integral limit and use those ones $t_1$ and $t_2$ given by Eq. (3.a) and (3.b).

**Remark 2: small values of r and v:** In the case of small r, Eq. (21) shows that $v_-$ tends to 0; so in this case the limits of integration of Eq. (21) are no more 0 and T, but $t_2$ and $t_1+T$ according to Fig. 1. These times $t_1$ and $t_2$ are given by Eqs. (3). Taking the limit of small r and of small $v_-/(b\omega)$, one expects that $t_1 << t_2$ , that $b\cos(\omega \Delta t)=v_- t_2$, with $T-t_2=\Delta t$; so, $t_1 \approx 0$ and $\Delta t \approx [4\pi v_-/(b\omega)]^{1/2}/\omega$ .

In order to solve the problem let us apply a self consistent argument and use the scheme of collisions proposed above; we will consider always that $t_1=0$, $T-t_2=\Delta t<<T$, $t_1<<\Delta t$ and $r<<1$. In a first step, we will consider the case of a typical particle subject to typical collisions to determine the typical scaling; then averaging will be performed to get the precise predicted behaviour at the leading order.

So, let us start with a typical particle with speed v; it hits the moving wall at time $t_2$ such as $v_-T= v_-(2\pi/\omega) \approx b[1-\cos(\omega t_2)]= b[1-\cos(\omega \Delta t)] \approx b\omega^2\Delta t^2/2$. This leads to:

$$v_-=b\omega^3\Delta t^2/(4\pi) \qquad (26.a)$$

Let bus now consider the collision with the moving wall; the typical time at which it occurs is $t_{coll}=T-\alpha\Delta t$ , with $\alpha < 1$; furthermore, since $v_-=v_+/r$, , the kinetic energy shall be mainly given by the wall, so that $v_-<<b\omega\sin(\omega\alpha\Delta t)$; the typical speed of the particle after the collision is then $v_+=b\omega(1+r)\sin(\omega\alpha\Delta t) =b(1+r)\alpha\omega^2\Delta t$ ; the particle then hits the fix wall and loose a great part of its kinetic energy since $r<<1$; so $v_-=r v_+= br(1+r) \alpha\omega^2\Delta t$. So:

$$v_-=rv_+=b\ r(1+r)\ \alpha\ \omega^2\ \Delta t. \qquad (26.b)$$

Compatibility of Eqs. (26.a) and (26.b) leads to the self consistent relation:

$$\omega\Delta t=4\pi\ \alpha\ r(1+r) \qquad (26.c)$$

Eq. (26.c) leads to the speeds:

$$V_+= v_+/(b\omega)= 4\pi\ \alpha^2\ r\ (1+r)^2 \qquad (27.a)$$

$$V_-= v_-/(b\omega)= 4\pi\ \alpha^2\ r^2\ (1+r)^2 \qquad (27.b)$$

This is then the predicted behaviour of $V_\pm$ at small r: $V_+$ shall scale linearly with r and $V_-$ as $r^2$ . This is indeed what will be observed in simulations. The coefficient $\alpha$ shall be computed; it is obtained from integration of Eq. (25) in between the limits $[-\Delta t,0]$ since $t_1\approx 0$. Then, when $v_-$ is small compared to $b\omega$, $\Delta t$ is defined by $v_-T\approx b[1-\cos(\omega \Delta t)]$ , which leads to Eq. (26.a); furthermore, as one can write $v_-T=\int_{-\Delta t}^{0} b\omega\sin(\omega t)dt=b[1-\cos(\omega t)]$, integration of Eq. (25.a) leads to:

$$<v_->= [r(1+r)(b^2\omega^4\ \Delta t^3)/3] / [b\omega^2\Delta t^2/2]= 2r(1+r)(b\omega^2\ \Delta t)/3 \qquad (28)$$

So, comparing Eq. (28) to Eq. (26.b) leads to $\alpha=2/3$. The values of the speeds predicted with Eq. (27) and $\alpha=2/3$ are reported in Table 2 for $r<0.2$.

***Remark 3: Study of the stability of the "equilibrium" solution:*** One can determine the stability of the predicted mean dynamics following the same guidelines as those ones established in §-1.2. It is enough to study the convergence of one speed, since the two speeds $v_-$ & $v_+$ are related together, *i.e.* $<v_->=r<v_+>$. So, writing $(v_{-eq})^2 = b^2\omega^2 r/[2(1-r)]$ which is the "equilibrium" solution, *i.e.* Eq. (24.c), and perturbing this solution, *i.e.* $<v_->= v_{-eq} +\delta v$, one gets by applying Eq. (25.a) that the mean speed $<v'_->= v_{-eq} +\delta v'$ after a round trip is such as:

$$\delta v' = \delta v(2r^2-1) \tag{29}$$

Since $0<r<1$, $|2r^2-1|$ is always smaller than 1; thus, the dynamics converges always towards $v_{-eq}$ and the nearer r is from $r_0=1/\sqrt{2}\approx 0.707$, the faster the return to equilibrium. Furthermore, there is a qualitative change of the dynamics of return to equilibrium when r crosses the value $r_o$: the dynamics converges and oscillates around the mean dynamics when $r<r_o$; it converges without oscillation when $r>r_o$. When r is too small or too large, *i.e.* $r\approx 0$ or $r\approx 1$, one gets $|2r^2-1|\approx 1$; it means that the convergence is slow; in this case, one shall also take into account the real distribution and calculate its broadening or sharpening after a round-trip to study the convergence; so, the real problem of the convergence is more complex in this case. At last, more caution shall be taken when $v_{-eq} <b\omega$, since the above theory shall be modified to take into account the fact that the limits of integration of Eq. (25) are no more [0, T] but are [$t_2$, $t_1$+T] and depends on $v_{-eq}$ according to Eq. (3). This modification will not be treated here.

### *1.4. RPA self consistent approach: case of 2 moving walls with different phases:*

As the previous approach is based on an averaging over times $\Delta t$ much longer than the period T, but smaller than $L/<v>$, it does not take into account any phase memory between two collisions (RPA hypothesis). It means that the results are independent of the phase difference between the two moving walls. It means that Eq. (14) describes also the motion of a bead in a box closed by two walls moving periodically with the same amplitude whatever the phase difference between the two wall motions. Thus, Table 1 gives also the mean speed variations as a function of the restitution coefficient in this more general case.

### *1.5. RPA Self consistent approach: The case of 2 moving walls with different phases and different amplitudes:*

RPA approximation supposes that the phase of the wall motion during the previous collision is forgotten when next collision arrives. So, using similar development as those used in previous subsections, using similar assumptions, and taking into account the amplitude of the wall motions, which we will call $b_1$ and $b_2$, the calculation leads to the self consistent conditions, when $<v> >b\omega$:

$$N_+ <v_+> = N_- <v_-> \tag{30.a}$$

$$N_+ <v_+>^2 = r\, N_- <v_->^2 + (1+r)(b_+\omega)^2\, N_-/2 \tag{30.b}$$

And due to symmetry and to the fact that the other wall has an amplitude b=0:

$$N_- <v_->^2 = r\, N_+ <v_+>^2 + (1+r)(b_-\omega)^2 N_+/2 \qquad (30.c)$$

## 2. 1 bead in a vibrating container: numerical simulation

Simulations have been performed to study the 1d dynamics of a single ball in a vibrating 1d container closed either by two walls performing sinusoidal motion of amplitude b in phase or in opposite phase or by a fix wall and by a second wall which vibrates. Let Ox be the unique direction of motion and x the coordinate of the particle; the particle bounces on the wall with a restitution coefficient defined as $r=v_{out}/v_{in}$, which is constant for a given simulation; no gravity is applied.

So since the wall(s) is (are) moving periodically, the domain which is accessible to the particle varies periodically with time: x pertains to [-L/2+f(t),L/2+g(t)], with f(t)=b cos($\omega$t) and g(t)=a f(t) where a=0, ±1 depending on the chosen boundary conditions (a=0 if the wall does not move, a=1 if it moves in phase with the other wall, a=-1 if it moves in opposite phase). The law of bouncing is given by :

$$v_{out}=(1+r)v_w - r\, v_{in} \qquad (31)$$

In Eq. (15), $v_w$ is the speed of the considered boundary at the time of collision $t_{coll}$:

$$v_w= df/dt= -b\omega \sin(\omega t_{coll}) \quad \text{or} \quad v_w= dg/dt= -ab\omega \sin(\omega t_{coll}) \qquad (32)$$

Using dimensionless argument, one expects that the dynamics of the particle shall be characterised by the distribution of the relative speed $V_i=v_i/(b\omega)$; it shall depend on the parameter set (a , b/L , r) and on the initial conditions ($v_o$, $\phi_o$)

So, for a given set of parameters (a, b/L , r) and for a given initial condition ($V_o$=2,$\phi_o$), the dynamics of the particle has been numerically calculated along a series of 2N successive collisions; influence of $\phi_o$ and of $v_o$ have been determined by computing the same series of collisions for 10 different initial phases: $\phi_i = \phi_o+\delta\phi i/10$, and 10 different initial speeds, with i=1 to 10; the effect of a very small $\delta\phi$, *i.e.* $\delta\phi<10^{-4}$ or of a rather large, *i.e.* $\delta\phi=2\pi/10$, has been investigated; in the same way, the range of investigated initial speed $v_o$ was $b\omega/100<v_o<2b\omega$. The relative mean speed $<V_\pm>$ in the +x and in the -x directions have been determined separately for each initial condition in order to test the importance of the memory effect. The standard deviation $\Delta V/<V>$ of $V_+$ or $V_-$ for each trajectory has been determined too . In some cases, it was observed that $V_i$, was depending on the initial phase, even for series of 1000 bounces; it means that the system does not exhibit some kind of ergodicity and has been keeping in memory the initial condition for a very long time. Such a situation is often observed for large b/L values. No more significant change of the results was observed when changing $v_o$; this is why the results concerning the $v_o$ dependence are not reported.

### *2.1. Variations of <V> as a function of* r *at fixed* b/L *and fixed* f*:*

This subsection discusses the results of simulations obtained with two walls vibrating in phase. Figs. 2-6 report the variations of the mean velocity $<V_{\pm}>=<v_{\pm}>/(b\omega)$ in both directions and of the relative standard deviation $\Delta V/V$ for different values of b/L, ranging from $10^{-3}$ to 0.1 as a function of the restitution coefficient r. As $<V_{\pm}>$ varies greatly as a function of r, the left part of the Figures reports $V_{\pm}$ values for small r (r<0.8) and the right one the $V_{\pm}$ values for larger r (r>0.8). In all these Figures the data from 10 different initial conditions are superimposed. One observes in all these Figs 2-6 that the data from $<V_{+}>$ and from $<V_{-}>$ are symmetric. This is due to the symmetry of the motion. One observes also that $\Delta V/V$ is smaller than 0.3 in general, except in some case where it can be even smaller, *cf*. Figs. 5 & 6.

Fig. 2 shows data obtained after averaging over 100 collisions and 5 000 collisions; it demonstrates that averaging becomes better and better, as $1/\sqrt{N}$, as a function of the number N of collisions in a given series, exactly as if the collisions were independent from one another in a given series. However, one observes also that fluctuations are much larger when r is in the range 0.95<r<1. This implies a large increase of memory effect.

This effect is more visible for b/L=0.005 or 0.01 for which resonant values of r are observed. It means that the average dynamics may depend on the initial condition for large r .

Comparison of Fig. 2 to Fig. 3 shows that the variation of <V> *vs*. r is independent of r at small b/L. Furthermore, these variations agree with the predicted values from the RPA approximation of §-1.2 (see Table 1). However, some discrepancy with this theoretical prediction is observed for large r values (r>0.97 for b/L=$10^{-3}$) .

Simulations for b/L=0.01 are reported in Fig. 4; they compare well also with the prediction from RPA at small r; but discrepancy becomes larger at large r due to some non ergodicity and to the apparition of some resonance at some r values, and for some initial condition.

Characteristics of the dynamics of the particle become quite different for b/L>0.05, which are reported in Figs. 5 & 6. It implies that the RPA approximation is no more valid in this case; this is normal since RPA assumes that b/L<<1; this gives then the range of b/L which is acceptable for RPA. In the case of large b/L, *i.e.* b/L>0.5, one observes that the mean particle speed is rather independent from r in a given range and then jump suddenly to an other value at a given threshold of r. For instance, in the case of b/L=0.05, $<V_{\pm}>$ is equal to 1 in the whole range of r from 0 to 0.4; then it jumps to $<V_{\pm}>\approx 2$ in the range 0.4<r<0.9 and to $<V_{\pm}>\approx 6$ in the range 0.92<r<1. In the case of b/L=0.1, the variations are more complicated; however they may be summed up as $<V_{\pm}>\approx 1-1.5$ in the range 0<r<0.6 and $<V_{\pm}>\approx 3.5-4$ in the range 0.6<r<1.

In the same way, one observes that the standard deviation $\Delta V/V$ for b/L=0.1 (Fig. 6) is often much smaller than in the previous cases (b/L<0.01), since it can be very close to 0.

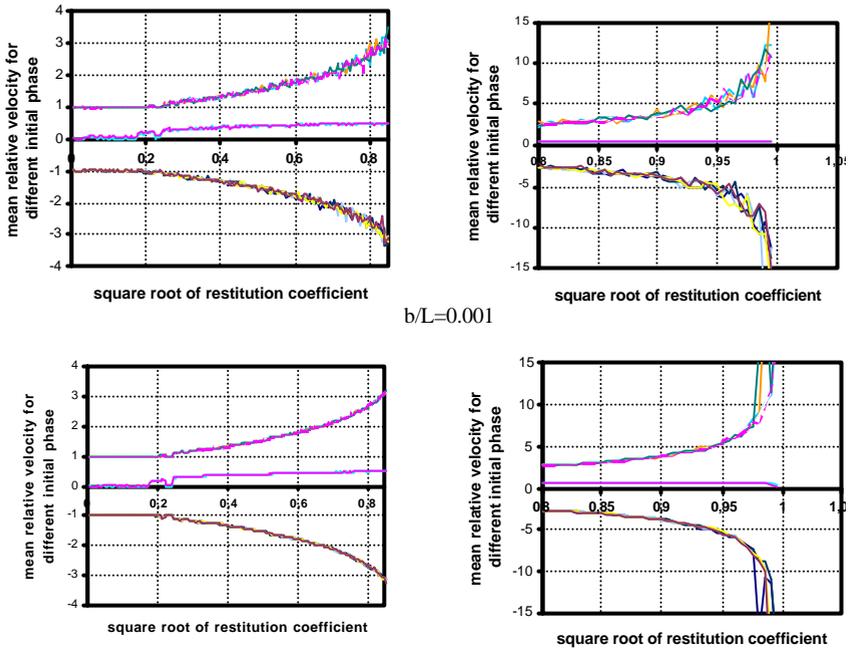

**Figure 2: 1 bead in a vibrating box:** *Variation of the mean relative speed* $V_\pm$ *and of* $\Delta V/V_\pm$ *as a function of the restitution coefficient* $r=(\varepsilon_E)^{1/2}$ *for 10 different initial phases and for an initial relative speed* $V_i=v_i/(b\omega)=2$; b/L=0.001; **upper figures:** *averaging over 100 collisions;* **lower figures:** *averaging over 5000 collisions;* f=30 Hz; *for curve legend see Figs.7 or 9.*

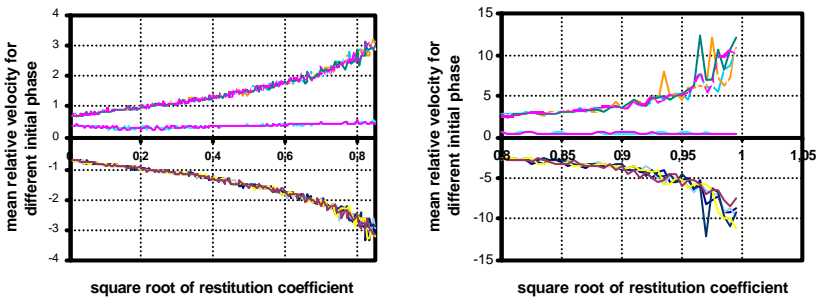

**Figure 3: 1 bead in a vibrating box:** *Variation of the mean relative speed* $V_\pm$ *and of* $\Delta V/V_\pm$ *as a function of the restitution coefficient* $r=(\varepsilon_E)^{1/2}$ *for 10 different initial phases and for an initial relative speed* $V_i=v_i/(b\omega)=2$; b/L=0.005; *averaging over 200 collisions;* f=30 Hz; *for curve legend see Figs.7 or 9.*

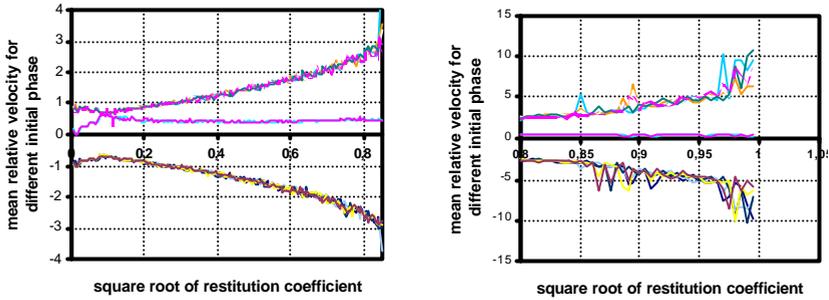

**Figure 4: 1 bead in a vibrating box:** *Variation of the relative mean speed $V_\pm$ and of $\Delta V/V_\pm$ as a function of the restitution coefficient $r=(\varepsilon_E)^{1/2}$ for 10 different initial phases and for an initial relative speed $V_i=2$; b/L=0.01; averaging over 200 collisions; f=30 Hz; for curve legend see Figs.7 or 9.* One sees that $\Delta v/v \to 0$ when $r \to 0$.

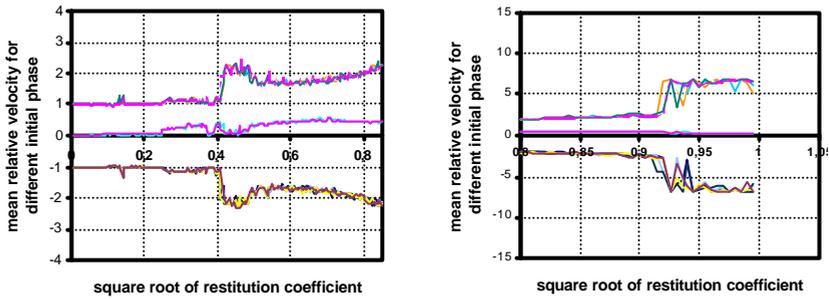

**Figure 5: 1 bead in a vibrating box:** *Same as Fig. 4 with b/L=0.05; averaging over 200 collisions; f=30 Hz; for curve legend see Figs.7 or 9.*

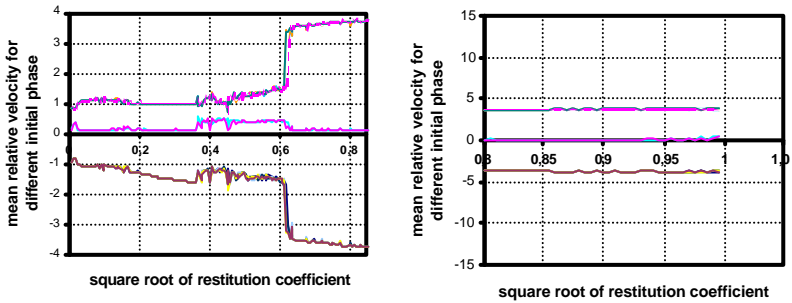

**Figure 6: 1 bead in a vibrating box:** *Same as Fig. 4 with b/L=0.1; averaging over 200 collisions; f=30 Hz; for curve legend see Figs.7 or 9.*

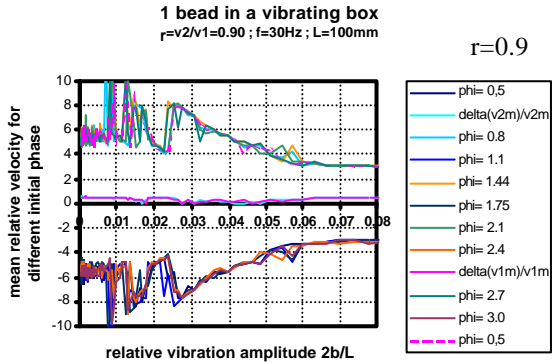

***Figure 7: 1 bead in a vibrating box:*** *Variation of the relative mean speed* $V_\pm$ *and of* $\Delta V/V_\pm$ *as a function of b/L for a restitution coefficient* $r=(\varepsilon_E)^{1/2}=0.9$ *for 10 different initial phases and for an initial relative speed* $V_i=v_i/(b\omega)=2;$ *f=30 Hz.*

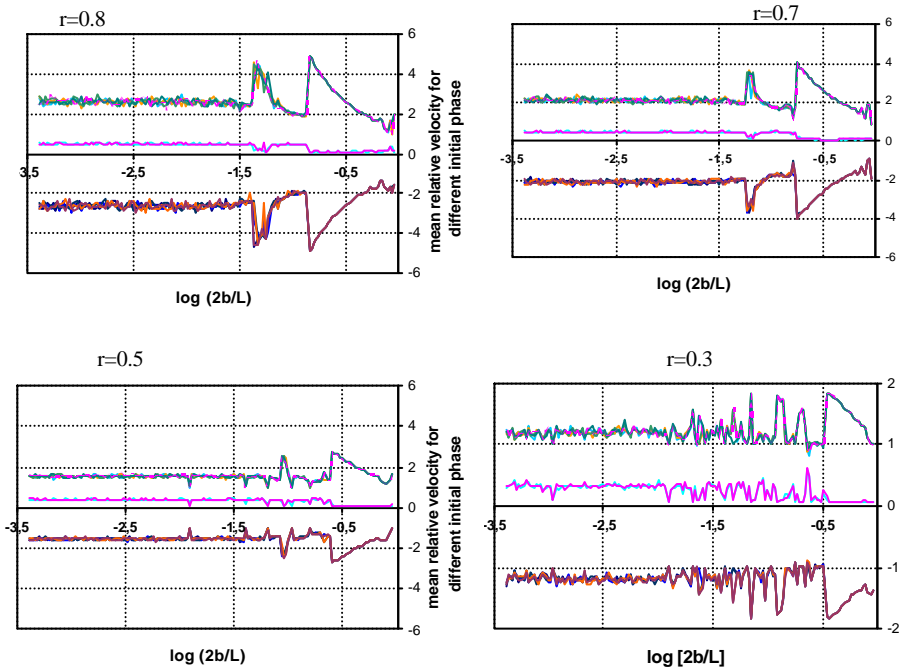

***Figure 8: 1 bead in a vibrating box:*** *Variation of the relative mean speed* $V_\pm$ *and of* $\Delta V/V_\pm$ *as a function of log(b/L) for 10 different initial phases and for an initial relative speed* $V_i=v_i/(b\omega)=2,$ *and for different restitution coefficient* $r=(\varepsilon_E)^{1/2}=0.8$, r=0.7, r=0.5, r=0.3; *f=30 Hz.; for curve legend see Figs.7 or 9.*

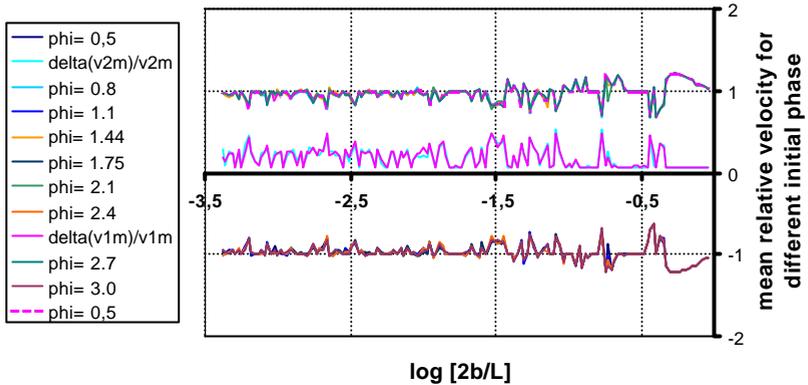

**Figure 9: 1 bead in a vibrating box:** *Variation of the relative mean speed $V_\pm$ and of $\Delta V/V_\pm$ as a function of $\log_{10}(b/L)$ for 10 different initial phases and for an initial relative speed $V_i = v_i/(b\omega) = 2$ for a restitution coefficient $r = (\varepsilon_E)^{1/2} = 0.1$; f=30 Hz. 10 different initial phases.*

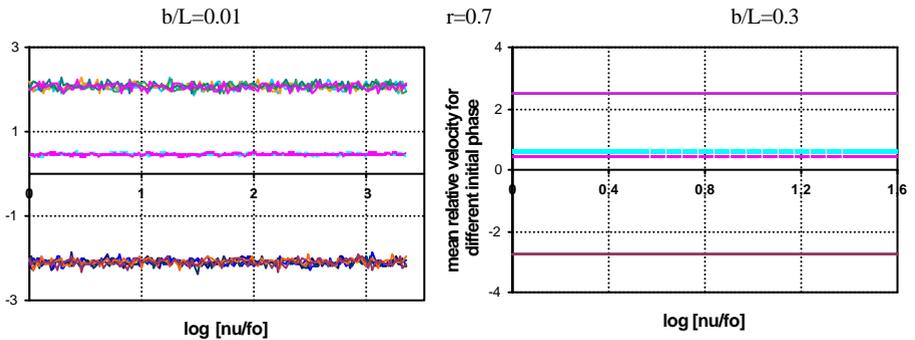

**Figure 10: 1 bead in a vibrating box:** *Two typical examples of the variation of the relative mean speed $V_\pm$ and of $\Delta V/V_\pm$ as a function of $f_o/f$, for 10 different initial phases and for an initial relative speed $V_i = v_i/(b\omega) = 2$; restitution coefficient $r = (\varepsilon_E)^{1/2} = 0.7$; $f_o = 30$ Hz.; for curve legend see Figs.7 or 9. This comparison has been performed for more different b/L ratios and for different r. All results show that $<V>$ and $\Delta V/V$ do not depend on $f_o/f$.*

## 2.2. Variations of <V> as a function of b/L at fixed f and fixed r:

Figs. (7-9) report the variations of the mean relative speed $<V> = v/(b\omega)$ and of the relative standard deviation $\Delta V/V$ as a function of $(2b/L)$ for a fixed frequency f=30Hz but at different restitution coefficient r, *i.e.* r=0.9, 0.8, 0.7, 0.5, 0.3 and 0.1. Logarithm scale is used for 2b/L in Figs. (8) & (9). $<V>$ and $\Delta V/V$ should be constant for a given r if the RPA approximation was holding true whatever b/L. Indeed, Figs (7-8) show that RPA approximation is only valid at very small b/L when r is larger than 0.1; Fig.

(9) shows that the RPA approximation is rather well satisfied at small r, *i.e.* r=0.1, even if one can still observe some b/L dependence at large b/L.

One should also remark that the relative variations of <V> as a function of 2b/L are larger for r=0.9 , *cf.* Fig. 7, since it can reach a factor of 3; these variations decrease when r decreases so that the factor is 2 for r ranging from 0.8 to 0.3. One can remark also that the amplitude of <V> variation decreases continuously with 2b/L when 2b/L is larger than 0.2 to 0.3, (or b/L>0.1 or 0.15); however, the precise value of 2b/L at which this decay starts depends on r, since it decreases when r increases from r=0.3 to 0.8 , *cf.* Fig. 8. Furthermore, before the constant decrease of <V>, one observes a transient regime of <V> *vs.* 2b/L; this transient can be constituted by a series of rapid variations, when r=0.9, 0.8, or by an other continuous decrease of <V>.

Anyhow, Figs. (7-9) demonstrate that RPA is always valid for small values of b/L; however the precise ratio depends on r; for instance it is b/L<0.005 for r=0.9, or b/L<0.02 for r=0.8, or b/L=0.25 for r=0.7… One can conclude from these simulations that this RPA approximation is always valid as soon as b/L>0.005. Indeed, this is in agreement with the finding of previous subsection.

At last, one shall remark that the relative standard deviation $\Delta V/V$ diminishes often when the system response <V> deviates from the RPA response. It means probably that a resonance is forced.

## *2.3. Variations of <V> as a function of* **f** *at fixed* **b/L** *and fixed* **r***:*

Fig. 10 reports variations of the mean relative speed as a function of the frequency of excitation f in units of $f_o$=30Hz, obtained from simulations of the dynamics of a single particle in a vibrating container for r=0.7 and two b/L values. These results show that $<V_\pm>=<v_\pm>/(b\omega)$ and $\Delta V/V$ are independent of f. Indeed more simulations have been performed; they concern two different values of the restitution coefficient and three different values of b/L; they all confirm that the dynamics of a single particle is independent of the excitation frequency. Indeed, this is normal, since the only time scale of the problem is 1/f as far as r is independent of the speed and as far as the particle is not submitted to gravity.

We have also observed that fluctuations of <V> diminish normally with the number N of collisions in a series, following a $\sqrt{N}$ law.

## *2.4. Simulations with a fix box with two walls vibrating in opposite directions:*

Simulations of the 1d dynamics of a single particle in a container closed by two walls vibrating with the same amplitude but in opposite directions have been performed. They give results similar to those ones found in subsections §-2.1 to §-2.3. It means that
- random phase approximation is valid when b/L<0.005 for r<0.95, and the <V> *vs.* r variations is the one given by Figs. 2 & 3
- that "non ergodicity" is found for large r and for b/L>0.005,
- that <V> varies by step at different values of r for b/L>0.05 and large r, *i.e.* when r>0.4;

- that fluctuations average normally as a function of the number of events
- that typical value of the relative standard deviation $\Delta V/V$ is still 0.3.
- that $<V>$ and $\Delta V/V$ do not depends on the frequency f.

## 2.5. Simulations of a bead in a fix box with one vibrating wall:

Simulations of the 1d dynamics of a single bead in a box closed by one fix wall and by one vibrating wall have been performed. The results have demonstrated that the relative mean speed $<V>=<v>/b\omega$ depends on which is the last collision; so $<v_+>/b\omega$ is different from $<v_->/b\omega$ ; it has been found also that $<V>$ and $\Delta V/V$ do not depends on frequency f. The typical relative standard deviation is still about 0.3 as in previous results. All this is normal and consistent with previous results.

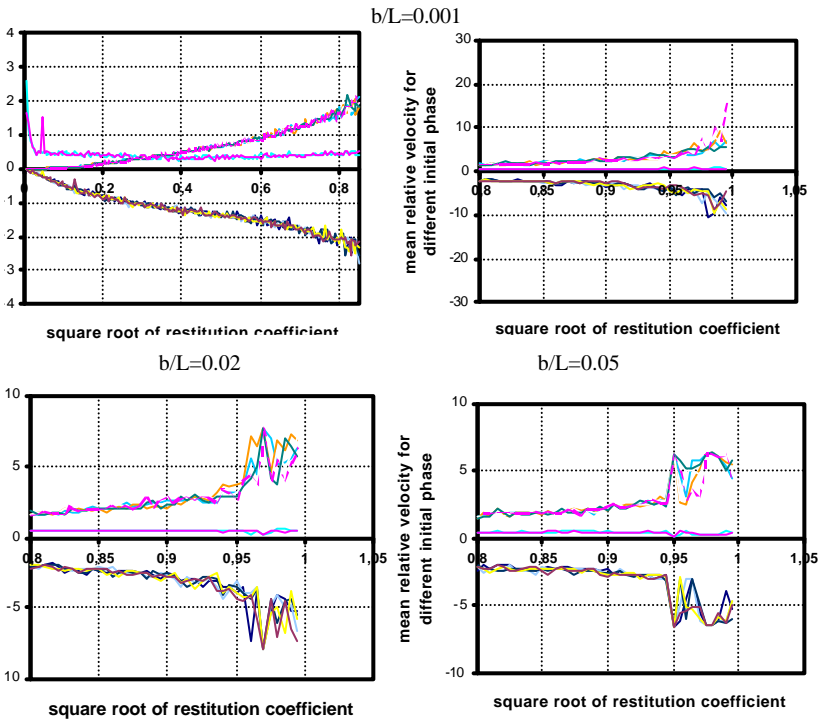

*Figure 11: 1 bead in a fix box with a vibrating wall:* Variation of the mean speed $V_\pm$ and of $\Delta V/V$ as a function of r , for different values of b/L and for 10 different initial phases and for an initial relative speed $V_i=v_i/(b\omega)=2$ : b/L=0.001 top figure, bottom-left: b/L=0.02, bottom right: b/L=0.05 ; frequency f=30 Hz.; for curve legend see Figs.7 or 9. Resonance at given values of r are observed for r>0.95 and b/L>0.02; non ergodic behaviour is also observed for r>0.95 and b/L>0.02 since averaged behaviour depends on the initial impact.

These results have shown also:

**r** *dependence of <V> at constant* **b/L***:* Figs. 11 & 12 report the variation of $<V_+>$ and $<V_->$ as a function of r for different values of b/L. One observes also that $<V_+>\neq<V_->$ for small values of r and the difference between the two speeds diminish when r increases; this is normal since the theoretical analysis predicts $<V_+>=r<V_->$ according to the geometry of the problem.

Comparison with Table 2 values shows that RPA model gives the correct dependencies of $V_\pm$ at small and at large r; it gives also the observed linear r dependence of $V_+$ at small r, with the right slope. Exact prediction in the range 0.15<r<0.5 would require the complete integration of Eq. (25) with no approximation. But data compare rather well with estimated values in this range too.

When b/L ⩾ 0.1 one observes step variations of <V> as a function of r, as in simulations using the other vibrated boxes. However, amplitudes of <V> are smaller. This is probably due to the fact that excitation is less important since only one wall is moving.

One observes also that the mean speed is much smaller than the one obtained when the two walls vibrate. This is due to the loss increase at a given wall motion.

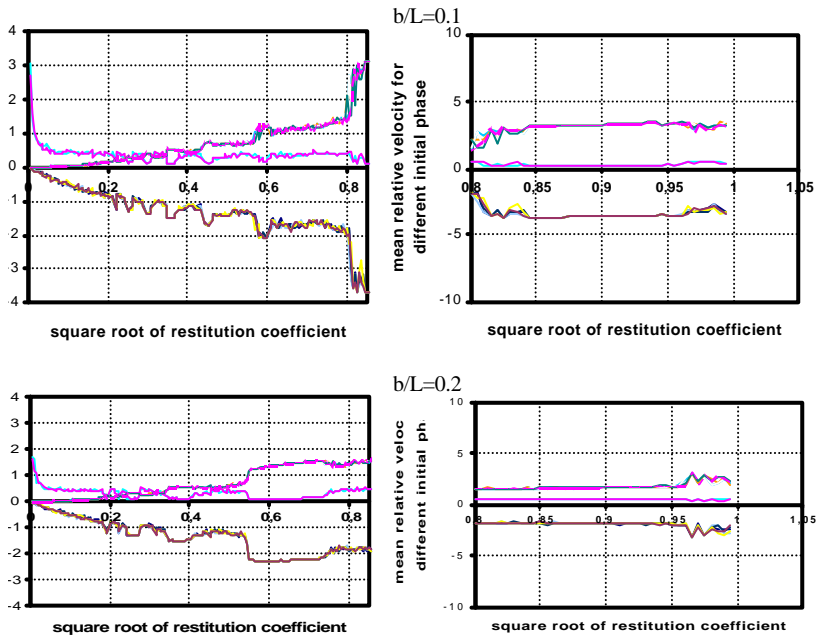

**Figure 12: 1 bead in a fix box with a vibrating wall:** *Variation of the mean speed $<V_\pm>$ and of ΔV/V as a function of r, for b/L =0.1 for 10 different initial phases and for an initial relative speed $V_i=v_i/(b\omega)=2$;(top figure) or b/L=0.2 ( bottom figure) ; frequency f=30 Hz.; for curve legend see Figs.7 or 9. $<V_\pm>$ varies by step. The behaviour looks rather always ergodic since the behaviour does not depend on the initial phase.*

***b/L dependence of <V> at constant r:*** Fig. 13 reports variations of $<V_+>$ and $<V_->$ as functions of $\log_{10}(2b/L)$ for different values of r. Resonant effect can be observed for r=0.3 for precise values of 2b/L>0.005; however these resonances correspond to small variations $\delta V$ of V, since $\delta V/V=0.4$. Anyhow, it means that the dynamics depends on b/L as soon as b/L>0.005.

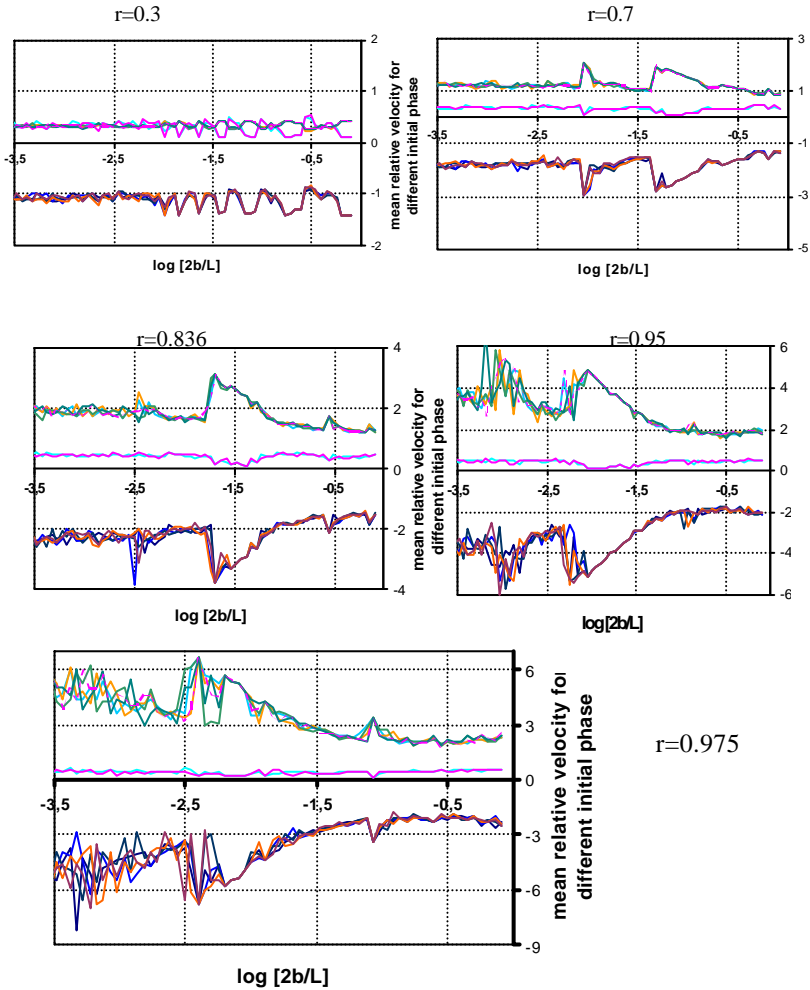

*Figure 13: 1 bead in a fix box with a vibrating wall: Variation of the mean speed $<V_\pm>$ and of $\Delta V/V$ as a function of b/L =0.1 for different values of r and for 10 different initial phases and for an initial relative speed $V_i=v_i/(b\omega)=2$: top left r=0.3, top-right: r=0.7, middle-left: r=0.836, middle-right: r=0.949, bottom: r=0.975 ; frequency f=30 Hz.; for curve legend see Figs.7 or 9. $<V_\pm>$ varies with b/L when b/L>0.005; when r is large, it decreases when b/L increases at large b/L (b/L>0.01). The behaviour looks rather ergodic except for some range of b/L and when r is large r.*

For larger values of r, *i.e.* r>0.7, one observes that <V> varies with b/L as it was also observed in the previous cases when the two walls of the container was vibrating; this b/L dependence occurs as soon as b/L>0.005 too: for instance, <V> decreases continuously when b/L increases at large values of b/L. The variation amplitude δ<V>/<V> is 1 about and this phenomena occurs in a range of b/L which depends on r: the domain of b/L is [0.02,0.2] if r=0.7, or [0.008,0.08] if r=0.84, or [0.002,0.03] if r=0.95, or [0.0015,0.03] if r=0.975 . The decrease of V is approximately linear in log(2b/L) scale, where log means $\log_{10}$, *i.e.* the logarithm in basis 10. One can also observe in Fig. 13 some resonance at smaller L/b values and for r=0.975.

## 3. Discussion and conclusion

The 1d dynamics of a single particle contained in a 1d box of size L has been investigated when the wall motion amplitude b is smaller than L/2 and for different boundary conditions which are (i) a vibrating box, (ii) one vibrating wall the other one being fix, (iii) two walls vibrating in opposite directions with the same amplitude. Theoretical modelling has been proposed using the random phase approximation (RPA) which assumes that the collision on the opposite wall does not remember the exact condition of the previous collision; theoretical demonstration of the validity of this approximation has been given for b/L ratios much smaller than 1. It has been found in this case that the mean relative speed V=v/(2πbf)=v/(bω) of the particle shall not depend on b and f, but does depend only on the collision restitution coefficient r=$v_{out}/v_{in}$.

Numerical simulations have been performed for different r, and at small b/L ratio, *cf.* Figs. 2 & 3; they have demonstrated that this RPA approximation works well when b/L is small enough , *i.e.* when b/L<<0.005. It means that the **"thermodynamics"** of a single bead in a container can be defined, in this case at least. This thermodynamics limit is characterised by two speed $v_{\pm}$ in the two directions, and by their distributions; it has been found their means <$v_{\pm}$> and their relative standard deviation δ$v_{\pm}$/$v_{\pm}$ depend on the speeds of the boundary conditions, but not on their relative phase. Moreover, as δ$v_{\pm}$/$v_{\pm}$ =0.2-0.3, it means that the $v_{\pm}$ =0 is quite unlikely, so that the whole speed distribution, formed by the distributions of $v_+$ and of $v_-$ present a dip in its centre, *i.e.* at v =0. So, owing to this, it is better to use the relative standard deviation of $v_+$ or of $v_-$ , *i.e.* δ$v_{\pm}$/$v_{\pm}$ , to define the temperature of the particle, rather than the total standard deviation δv since δv=($v_+$-$v_-$)/2≈<$v_+$>≈<$v_-$> in the present case. The relative mean velocity <$v_{\pm}$>/(bω) is a function of the unique parameter r, *i.e.* the restitution coefficient; this function depends on the boundary conditions, but does not depend on f.

In the same way, one can define the **thermodynamics of a collection of non interacting particles** which are contained in a 1d vibrating box; these particles are supposed to interact with the box walls, but do not interact together. Within such an hypothesis, this thermodynamics is the one obtained from RPA model. It means in particular that one can define a pressure which will be the mean transfer of momentum due to collisions of the beads with the walls. This pressure will be proportional to the density of the particles and to the momentum transfer. It shall scale as $(b\omega)^2$ for a given restitution coefficient; however, the pressure measured on the wall will vary periodically with the wall velocity, due to the variation of the wall velocity.

   Numerical simulations performed at small and large b/L and at different frequencies have shown that the renormalised velocity $v_{\pm}/(b\omega)$ is always independent of f (or of $\omega=2\pi f$), whatever the range of b/L<0.5, *cf.* Figs 9 & 10. This is in agreement with simple theoretical argument, since there is no specific physical time scale in the problem, except the one imposed by the vibration frequency. The problem would have been probably different if the two walls were vibrating at two different frequencies; in this case, the $\omega_1/\omega_2$ ratio should be important at large b/L .

   Random phase approximation (RPA) becomes not valid in some cases when b/L ratio becomes larger than 0.005. However, in most cases, when r is smaller than 0.9-0.95 and b/L not too large, but b/L>0.005, it seems that ergodicity is still observed, which means that the thermodynamics of the isolated particle in the vibrating container or of a gas of non interacting particle can still be defined; but this thermodynamics depends not only on $b\omega$ but also on b/L. This has been checked by looking at the statistical distribution of the $v_{\pm}$ for different initial conditions, changing the initial phase by a very tiny amount or by a large one, *i.e.* $\delta\phi_o<10^{-4}$ or $2\pi/10$, or changing the initial speed in the range $b\omega/1000<v_o<2b\omega$ . However, the variation of the mean velocity $<v_{\pm}>$ as a function of r deviates from the one computed by the RPA approximation in this range of b/L. In particular, the $<v_{\pm}>/(b\omega)$ varies rather by steps as a function of r at a given b/L as soon as b/L>0.05, *cf.* Figs. 5 & 6. Furthermore, this behaviour is associated to a specific type of variation of the mean relative speed $<v_{\pm}>/(b\omega)$ as a function of b: one observes at a fixed r that $<v_{\pm}>/(b\omega)$ jumps suddenly for a specific b/L value, then it decreases continuously as 1/b approximately as it can be observed in Figs 7 & 8. It means that the height of the step in $<v_{\pm}>/(b\omega)$ decreases with the increase of b/L. These points are characteristics of the thermodynamics at large b/L ratio.

   At last Figs. 8 & 9 shows some resonances for specific values of b/L when r is small, *i.e.* r=0.3 and r=0.1. This may complicate the true statistical behaviour of the ball dynamics.

   In the same way, resonance effects become more obvious at larger values of r, *i.e.* when 0.85-0.95<r<1, as soon as b/L overpasses b/L=0.005. These resonances are characterised by (i) larger values of $<v_{\pm}>/(b\omega)$ for some values of the phase $\phi$, (ii) increase of the sensitivity of $<v_{\pm}>/(b\omega)$ on the initial condition, (iii) decrease of the relative standard deviation $\delta v/<v_{\pm}>$, and/or (iv) loss of ergodicity. All these effects are not always observed at the same time. However, they do indicate some resonance and some long memory effect, provoking probably the breaking of ergodicity in some cases. At last, the effect described in the last paragraph and which corresponds to Figs. 7 & 8 have to be included in this discussion so that the leading effect from an experimental point of view is most likely the decrease of the relative mean velocity $<V>=v/(b\omega)$ when b/L is increased in the range 0.005-0.1. The other leading effect is probably the loss of ergodicity so that the mean dynamics shall be obtained in a finite time after averaging over all the initial conditions at large r, *i.e.* r>0.95, *cf.* Fig. 3.

It is also found that the relative standard deviation of the velocity $\Delta V/V$ which corresponds to a single trajectory is always in the range 0.2 — 0.4; it means that it does not vary, except for specific case of resonance as those ones reported in Fig. 9.

The simulations have demonstrated that non ergodicity occurs via resonant effect at large r and b/L values . Such breaking of ergodicity occurs most likely due to the selection of a discrete set of possible phases at which collisions can occur with walls. In this case one can observe that the mean speed of the particle, obtained after averaging over a large number of collisions $N_t$=100 or more, does depend on the initial condition.

This paper has shown also that the mean speed of the particle is smaller when the particle is excited via a single moving boundary than via the two, keeping constant the amplitude b. It has shown also that the mean speed $(<|V_+|>+<|V_-|>)/2$ expressed in vibration unit $b\omega$, *i.e.* $<V>=<v>/(b\omega)$ shall be larger than 1 for restitution coefficient r larger than a given value $r_o$ which depends on the set-up, *i.e.* $r_o \approx 0.5$ for 1 vibrating wall, $r_o$=0.2 for 2 vibrating walls.

In the same way, the mean roundtrip time $T_r$ depends strongly on r and on the experimental configuration; it is not strictly proportional to $1/<v_+>$ or $1/<v_->$ since $T_r = <(L/v_+ + L/v_-)>$; but as $\delta v_\pm/v_\pm$ is small, it is given approximately by $T_r=\{L/<v_+>+L/<v_->\}$. So $T_r$ is approximately $2L/<v>$ when the two walls vibrate, but it is approximately $(r+1)L/<v_+>=(1+1/r)L/<v_->$ when only one wall vibrates since $v_+=rv_-$ in this case.

At last it is worth strengthening that simulations do show that the mean relative speed $V=v/(b\omega)=v/(2b\pi f)$ decreases continuously when b/L increases in the range 0.01<b/L<0.2. But the amplitude of the decrease diminishes when r decreases; this amplitude of variation can be as large as a factor of 2 or 3. This may be important since b/L ranges within this domain in most experimental cases, because it allows to visualised at the same time the particle dynamics and the wall motion. In particular, this is the case in the rocket experiments reported in [4,5]. So the results reported in this paper let think that under these experimental conditions, *i.e.* r large =0.5 and b/L large, the very low density "gas" shall exhibit an anomalous behaviour as a function of b. This may explain perhaps the anomalous dependence $<v>(b\omega)^{3/2}$ reported in [4,5].

So, we hope this work helps understanding experiments on "granular gas" in low gravity conditions [4,5] since the gas state which has been observed seem to be in a Knudsen regime. This will be the topics of a forthcoming article.

*Acknowledgements:* CNES is thanked for funding.

# Feed-back from Readers :

## *Discussion, Comments and Answers*

## From *poudres & grains* articles:

On *Poudres & grains* **12**, 17-42 (2001): In Figs 8,9,10, l3, Log means Neperian Logarithm. L/2 is the half of box length in Figs. 11, 12 & 13.
  *From a remark by Y. Garrabos:* The rapid decrease of $V/(b\omega)$ *vs.* $\ln(2b/L)$, with the increase of b/L, is due to the synchronisation of the bead motion on the excitation, leading to regular impact at a frequency equal to half the excitation frequency $\nu$ {this occurs around $V/(b\omega)=4$ and $\ln(2b/L)=-2.53$, *i.e.* and corresponds to a one way ($\approx L$) which takes at time $1/\nu$} or at $\nu$ {this occurs around $V/(b\omega)=4$ and $\ln(2b/L)=-1.84$, *i.e.* round-trip ($\approx 2L$) takes $1/\nu$ }. This last synchronisation has been observed in recent Airbus result.

<div style="text-align: right">P.E.</div>

On *Poudres & grains* **12**, 60-82 (2001): *Remark by L. Ponson, P. Burban, H. Bellenger & P. Jean* : According to the data, $l_c=L/5$ to $L/6$ for the less dense sample.
  *Answer:* Indeed, the mean free path $l_c$ is smaller than the cell size L; however, it remains of the same order; so, this does not invalidate the qualitative findings and the theoretical discussion about the Knudsen regime; moreover, one expects that a decrease of $l_c/L$ generates an increase of loss, this value of $l_c/L$ may help understanding why the mean bead speed is so slow in the MniTexus experiment, compared to what one shall expect from 1-bead simulations.

<div style="text-align: right">P.E.</div>

On *Poudres & grains* **12**, 107-114 (2001): lines 6-7 of 2$^{nd}$ paragraph of section 3: the liquid flow exists but is small in the limit of a small toroidal section of radius $R_s$, *i.e.* proportional to $R_s/R_t$, due to preservation rules as stated in the introduction. In this limit this affects little the results. Furthermore, it is known that ripples formation occurs above the threshold at which the Stokes boundary layer is destabilised, *i.e.* when the Reynolds number $R_{e\delta}= b\Omega\delta/\nu=2b/\delta$ is larger than 100-to-500, for which $\delta=\sqrt{(2\nu/\Omega)}$ is the viscous boundary layer thickness and b is the relative amplitude of the flow motion; so, in the present case it scales as $R_{e\delta}=2\alpha_m R_s/\delta$, for which $\alpha_m$ is the amplitude of rotation. It has been found also that ripple wavelength $\lambda$ is about $12\delta$ at threshold. Limit of Stability of the Stokes boundary layer can be found in V.G. Kozlov, Stability of periodic motion of fluid in a planar channel // Fluid Dynamics, 1979, vol. 14, no. 6, 904–908 and in Stability of high-frequency oscillating flow in channels // Heat Transfer – Soviet Research, 1991, v. 23, no. 7, 968–976.

P.E.

# About published articles from other Reviews

About *Phys.Rev. Lett*. **84**, (2000) 5126, and *C.R. Physique* **3** (2002) 217-227, by J. Duran: Formation of such Ripples and its domain of existence has been studied under sinusoid vertical forcing by V.G. Kozlov, A. Ivanova and P. Evesque (Europhys. Lett. **42**,413-418 (1998)) with a viscous fluid instead of air. Similar heap formation has been found; it was also found .

P.E.

The electronic arXiv.org version of this paper has been settled during a stay at the Kavli Institute of Theoretical Physics of the University of California at Santa Barbara (KITP-UCSB), in june 2005, supported in part by the National Science Fundation under Grant n° PHY99-07949.

*Poudres & Grains* can be found at :
http://www.mssmat.ecp.fr/rubrique.php3?id_rubrique=402